# Time evolution of entropy associated with diffusivity fluctuations: Diffusing diffusivity approach

Yuichi Itto

*Science Division, Center for General Education, Aichi Institute of Technology,*

*Aichi 470-0392, Japan*

**Abstract**   It has experimentally been found by Lampo *et al*. [Biophys. J. **112**, 532 (2017)] that, for two different types of cell, the distribution of the diffusivities of RNA-protein particles over cytoplasm obeys an exponential law. Then, an interesting issue has been pointed out: this exponential distribution is the maximal entropy distribution. Here, time evolution of entropy associated with local fluctuations of the diffusivity is studied. The entropy rate under the diffusing diffusivity equation, which admits the exponential fluctuation as its stationary solution, is shown to be positive. The present result is expected to be useful for studying the dynamics of diffusivity fluctuations. Furthermore, the distribution of time being required for characteristic displacement of the RNA-protein particle is found to decay as a power law. A comment is also made on a formal analogy with the thermodynamic relation concerning temperature.

A recent experimental study in Ref. [1] has elucidated an exotic property of diffusion of RNA-protein particles in living cells. A messenger RNA fluorescently labeled with a protein is regarded as a particle. Such a property has been found for two different types of cell: one is *Escherichia coli* cell (i.e., a bacterium), and the other is *Saccharomyces cerevisiae* cell (i.e., a yeast). Trajectories of the RNA-protein particles in cytoplasm of each cell have been analyzed at the level of individual trajectory. Then, the mean square displacement that behaves for elapsed time, $t$, as

$$\overline{x^2} \sim D\, t^{\alpha} \qquad (1)$$

has been obtained, which characterizes the diffusion property of the particle. Here, $D$ stands for the diffusion coefficient, which is referred to as diffusivity, and $\alpha$ is termed the diffusion exponent.

The results of the analysis of experimental data are as follows. In comparison with the diffusion exponent that is obtained through the mean square displacement over the trajectories and takes a positive value less than unity, the values of $\alpha$ in Eq. (1) are seen to be approximately equal to this value at each cell type. Such a phenomenon is called anomalous diffusion [2], which is under vital investigation in the literature [3,4]. It originates from viscoelastic nature of the cytoplasm [1]. In contrast to the feature of $\alpha$, the following result about the diffusivity is intriguing. As can be seen in Fig. 3 in Ref. [1], $D$ fluctuates in a wide range, and its distribution well obeys the following exponential law:

$$P(D) \sim \exp(-D/D_0), \tag{2}$$

where $D_0$ is a certain positive constant having the dimension of diffusivity and gives a typical value of diffusivity. This distribution plays a crucial role for describing non-Gaussian distribution of the displacements of the RNA-protein particles.

Regarding the distribution in Eq. (2), a fact to be emphasized is its *robustness* in the following sense [1]: the distribution holds for two different types of cell (i.e., *Escherichia coli* and *Saccharomyces cerevisiae*). The trajectories of the RNA-protein particles are expected to be observed in different areas of the cytoplasm [1]. Thus, the results of the data analysis suggest the existence of universal feature of the local fluctuations of the diffusivity over the cytoplasm.

Here, it seems appropriate to make some comments on the importance of the exponential distribution of diffusivities. In an experimental study in Ref. [5], the distribution of the displacements of vesicles in a solution, which exhibit non-Gaussian normal diffusion, has been discussed based on the exponential distribution of diffusivities. Then, it has been shown in a recent experiment in Ref. [6] that non-Gaussian distribution of the displacements of proteins on a cell membrane can be related to dynamic heterogeneity of diffusivity, whose distribution is of the exponential form. In these studies as well as the work in Ref. [1], the distribution of the displacements is described by the superposition of the Gaussian distribution of the displacements with respect to the distribution of diffusivities. In this respect, it may be of interest to point out that such a description has been discussed from the perspective of the so-called superstatistics [7] in Ref. [8] (see Refs. [9,10] for analogous discussions in the context of the diffusion exponent).

The origin of diffusivity fluctuations reported in Ref. [1] remains unclear, although the distribution in Eq. (2) may imply its simplicity. Quite remarkably, however, the following issue, which should be addressed, has been pointed out [1]: the exponential distribution in Eq. (2) is the *maximal entropy* distribution.

In this paper, we discuss time evolution of entropy associated with the local fluctuations of diffusivity, $D$, in Eq. (1). For it, we first derive the exponential distribution of the fluctuations based on a maximum-entropy-principle approach, which has been proposed in Refs. [11,12] in the context of the diffusion exponent (see Ref. [13] for a discussion about the maximum entropy principle for deriving the distributions of fluctuations of inverse temperature in complex systems). Then, to study the time evolution, we employ the diffusing diffusivity equation developed in Ref. [14] (see Refs. [15,16] for recent developments, for example), which offers a description of the dynamical behavior of the fluctuation distribution and admits the exponential distribution as its stationary solution. We show how the entropy rate turns out to be positive. Thus, the result is expected to be useful for studying the dynamics of diffusivity fluctuations found in Ref. [1]. Furthermore, the distribution of time being required for characteristic displacement of the RNA-protein particle is found to decay as a power law, which has an explicit connection with the result obtained from an experiment similar to the one in Ref. [1]. Regarding the entropy, we also make a comment on a formal analogy with the thermodynamic relation concerning temperature.

First of all, we regard the cytoplasm of each of *Escherichia coli* and *Saccharomyces cerevisiae* cells as a medium for diffusion of the RNA-protein particle and imaginarily divide this medium into many small blocks. In each block, the particle exhibits anomalous diffusion with the form in Eq. (1). Note that the diffusivity $D$ fluctuates, depending on the blocks. Although the diffusivity may slowly vary in each block on a long time scale, we assume that $D$ is approximately constant. (This assumption will be relaxed in our later discussion.) We are considering a situation where the local property of the fluctuations of the diffusivity over the cytoplasm is unknown. In such a situation, we shall introduce a measure of uncertainty about the local fluctuations in order to study the statistical property of the fluctuations: that is, the entropy associated with diffusivity fluctuations. This entropy is related to information about how the diffusivity locally fluctuates over the cytoplasm in such a way that the larger the uncertainty is, the larger the entropy is. (Later, it turns out to take the form of the Shannon entropy.) We suppose that the medium consists of the blocks with a set of different values of the diffusivity, $\{D_i\}_i$ with $D_i$ being the $i$-th value of the diffusivity. Then, we construct replicas of the medium, each of which is distinct from each other with respect to the local property of diffusivity fluctuations. Note that the statistical property of the fluctuations in each replica is equivalent to that in the medium. We denote the number of replicas by $\Omega$, which is expected to be large since the medium contains many blocks. With this, we introduce the entropy associated with diffusivity fluctuations as follows:

$$S = \frac{\ln \Omega}{N}, \qquad (3)$$

where $N$ is the total number of blocks in the medium.

Clearly, we need to evaluate $\Omega$, otherwise Eq. (3) is formal. To do so, recall that the mean square displacement in Eq. (1), by which $D$ is determined, is obtained from individual trajectory. This means that a given trajectory is included in average taken in the analysis of $\overline{x^2}$ in Eq. (1) and other trajectories in different areas of the cytoplasm are not. This implies that the local blocks are independent each other in terms of the diffusivity. Accordingly, denoting the number of blocks with $D_i$ in the medium by $n_{D_i}$, which satisfies $\sum_i n_{D_i} = N$, $\Omega$ is given by $\Omega = N!/\prod_i n_{D_i}!$, leading to the following form of the Shannon entropy: $S \cong -\sum_i P_{D_i} \ln P_{D_i}$, where $P_{D_i} = n_{D_i}/N$ is the probability of finding $D_i$ in a given block of the medium, provided that $\ln(M!) \cong M \ln M - M$ for a large value $M$, which represents $N$ and $n_{D_i}$'s, (i.e., the Stirling approximation) has been employed.

Now, it seems necessary to examine how large the number of blocks is. Below, we shall see this by evaluating the volumes of both the cytoplasm and the block.

As a tractable case from the data analysis, we consider the case of *Saccharomyces cerevisiae* cell only. According to Ref. [1], the typical diameter of this cell is given by $3\,\mu\text{m}$. Assuming that the cell is a sphere with this diameter, the volume of the cell is estimated to be $14\,\mu\text{m}^3$, and accordingly, the volume of the cytoplasm is less than this volume (due to the presence of cell nucleus). Now, for a given local block, let us regard it as a cubic block, which has the value of $\sqrt{\overline{x^2}}$ at large elapsed time as the length of its side. In this way, it is possible to estimate the volume of the block. As an example of the value of $D$ in Eq. (1), we choose $D \sim 0.01\,\mu\text{m}^2/\text{s}^\alpha$, which is taken from Fig. 4 (c) in the Supplemental Information for Ref. [1]. Then, the value of the diffusion exponent is given by $\alpha = 0.75$ and the elapsed time is taken to be $t = 0.75\,\text{s}$ [1]. (Although this value of $\alpha$ is the one obtained through the mean square displacement over the trajectories, we employ it since the exponent to be used in $\overline{x^2}$ seems to be in the vicinity of the value [1].) Thus, in this example, the volume of the cubic block is estimated to be $0.0058\,\mu m^3$.

So, in the case when no additional blocks with the above-chosen value of $D$ exist, if the number of blocks with the above-estimated volume in the medium is, for instance, 100, then $n_{0.01} = 100$ follows. Correspondingly, the total volume of these blocks in the medium is given by $0.58\,\mu m^3$, which is supposed to be smaller than the volume of the cytoplasm. It is therefore implied that $n_{0.01}$ is actually larger than the value taken and the number of blocks with each of other different values of $D$ in the medium is also large. This comes from a natural requirement that the volume of the medium is equal to that of the cytoplasm. Thus, it is reasonable to suppose that the number of blocks with each value of the diffusivity in the medium is large enough, (allowing us to use the Stirling approximation).

We describe the continuum limit of the entropy as

$$S[P] = -\int dD\, P(D)\ln P(D), \tag{4}$$

where $P(D)dD$ is the probability of finding the diffusivity in the interval $[D, D+dD]$. It is noted that $P(D)$ has the dimension of $D^{-1}$ and a positive constant should be introduced inside the logarithm due to the dimensional reason. This turns out to bring an additive constant, and therefore $S[P]$ is determined up to such an additive constant. The same notation $P(D)$ as that in Eq. (2) is used for this probability density, but it may not cause confusion.

Let us derive the exponential distribution of the fluctuations of $D$ by the maximum entropy principle [17] associated with the entropy $S[P]$. As mentioned earlier, the situation we consider here is that there is no knowledge about the local property of such fluctuations. In this situation, suppose that we are given information about the average of $D$, the constraint on which is therefore imposed as

$$\int dD\, P(D)\, D = \overline{D}. \tag{5}$$

Together with another constraint on the normalization of $P(D)$, the entropy is maximized with respect to $P(D)$:

$$\delta_P \left\{ S[P] - \nu \left( \int dD\, P(D) - 1 \right) - \lambda \left( \int dD\, P(D)\, D - \overline{D} \right) \right\} = 0, \tag{6}$$

where $\nu$ and $\lambda$ are the Lagrange multipliers associated with the constraints on the normalization of $P(D)$ and the average of $D$, respectively, and $\delta_P$ stands for the variation with respect to $P(D)$. Now, it is natural to expect that the RNA-protein particles are hindered by elements in the cytoplasm and accordingly its diffusion is suppressed (i.e., not enhanced). In terms of the diffusivity, this may be realized as follows: no blocks with large $D$ exist. Therefore, we impose the following condition

$$\lim_{D \to \infty} P(D) = 0, \tag{7}$$

which turns out to make $\lambda$ positive. Consequently, we have the following stationary solution of Eq. (6): $\hat{P}(D) \propto \exp(-\lambda D)$. Thus, it is obvious that this distribution is equivalent to the exponential distribution in Eq. (2) after the identification, $\lambda = 1/D_0$, is made.

Up to this stage, the diffusivity is assumed to be approximately constant. In what follows, we relax this assumption. There, the fluctuation distribution of diffusivity is considered to vary in time and accordingly deviates slightly from the exponential distribution, in general. Such a distribution seems tend to move towards the exponential distribution. A question of interest here is: how does the entropy in Eq. (4) evolve in time? [Even for the fluctuation distribution that the entropy does not take its maximum value, the entropy in Eq. (4) seems to be useful, if the change of $\Omega$ from its maximum value brings a negligible quantity for the maximum value of the entropy itself due to the largeness of $N$ (see Ref. [18] for an analogous discussion).]

We here wish to answer this question by employing the diffusing diffusivity equation presented in Ref. [14], which has been proposed for obtaining non-Gaussian distribution of the displacements of particles exhibiting normal diffusion. This equation describes time evolution of the distribution of diffusivities, which obeys a diffusion equation of advection-diffusion type in terms of the diffusivity. Therefore, such a mechanism is called "diffusing diffusivity". The equation admits the exponential distribution as its stationary solution.

For the probability, $P(D, t)dD$, of finding the diffusivity in the interval $[D, D+dD]$ at time $t$ (i.e., the conventional time), let us consider the following diffusing diffusivity

equation [14]:

$$\frac{\partial P(D,t)}{\partial t} = -\frac{\partial J(D,t)}{\partial D}, \tag{8}$$

where $J(D,t)$ is the probability current given by

$$J(D,t) = -\frac{\partial}{\partial D}\bigl[k(D)P(D,t)\bigr] - s(D)P(D,t) \tag{9}$$

and satisfies the following reflecting boundary conditions

$$J(0,t) = 0, \qquad \lim_{D \to \infty} J(D,t) = 0. \tag{10}$$

Here, $k(D)\,(>0)$ stands for the diffusivity of the diffusivity, (for which different notation is used here), whereas $-s(D)\,(\leq 0)$ describes the bias of the diffusion of the diffusivity. The fluctuation distribution as well as its derivative with respect to the diffusivity are supposed to vanish in the limit, $D \to \infty$. This gives the latter condition in Eq. (10), and the former one comes from Eq. (8) under the normalization condition of the fluctuation distribution, accordingly. As shown in Ref. [14], in the case when $k(D)$ and $s(D)$ are certain positive constants, which are denoted respectively by $k$ and $s$, the stationary solution of Eq. (8) is given by the exponential distribution. This distribution with $D_0 = k/s$ is identical to the one in Eq. (2).

We are interested in how the following Shannon entropy

$$S(t) = -\int dD\, P(D,t) \ln P(D,t) \tag{11}$$

evolves in time under the diffusing diffusivity equation in Eq. (8). The time derivative of the entropy is written as follows:

$$\frac{dS}{dt} = \int dD \frac{[J(D,t)]^2}{k(D)P(D,t)} + \frac{d}{dt}\left\langle \frac{s(D)}{k(D)} D \right\rangle$$

$$+ \int dD \frac{J(D,t)}{k(D)} \left\{ \left[ \frac{s(D)}{k(D)} D + 1 \right] \frac{dk(D)}{dD} - D \frac{ds(D)}{dD} \right\}, \tag{12}$$

provided that Eqs. (8), (9), and the conditions in Eq. (10) have been used, and the symbol $\langle Q \rangle$ denotes the average with respect to the fluctuation distribution $P(D,t)$: $\langle Q \rangle \equiv \int dD P(D,t) Q$. The entropy rate is purposely expressed in terms of both the probability current and the average.

In the above-mentioned case when $k(D)$ and $s(D)$ are the positive constants satisfying $D_0 = k/s$, which is precisely the case we are interested in, Eq. (12) becomes simplified to be

$$\frac{dS}{dt} = \frac{1}{k} \int dD \frac{[J(D,t)]^2}{P(D,t)} + \frac{s}{k} \frac{d\langle D \rangle}{dt}. \tag{13}$$

This entropy rate is expected to be positive, since the fluctuation distribution approaches the exponential distribution in Eq. (2). From Eq. (13), this point is clarified by examining time evolution of $\langle D \rangle$, as we shall see below.

Using Eqs. (8), (9), and the conditions in Eq. (10), the time derivative of $\langle D \rangle$ is calculated to be

$$\frac{d\langle D \rangle}{dt} = k\, P(0,t) - s, \tag{14}$$

provided that $\lim_{D \to \infty} P(D,t) = 0$. To evaluate the right-hand side, it seems necessary to obtain the solution of Eq. (8). We suppose that the initial value of the diffusivity is taken to be zero, since it may be natural to consider that no trajectories of the RNA-protein particles over the cytoplasm exist at the initial time. Therefore, the solution of Eq. (8) satisfying the initial condition, $P(D,0) = \delta(D)$, is obtained as follows [19]:

$$P(D,t) = \frac{1}{\sqrt{\pi k t}} \exp\left[-\frac{(D+st)^2}{4kt}\right] + \frac{s}{k\sqrt{\pi}} e^{-\frac{sD}{k}} \int_{\frac{D-st}{2\sqrt{kt}}}^{\infty} dy\, \exp(-y^2). \tag{15}$$

Accordingly, Eq. (14) is found to be given by

$$\frac{d\langle D \rangle}{dt} = \frac{s}{2\sqrt{\pi}} \int_{\frac{s}{2}\sqrt{\frac{t}{k}}}^{\infty} dy\, y^{-2} \exp(-y^2), \tag{16}$$

showing that $d\langle D \rangle / dt > 0$, i.e., $\langle D \rangle$ monotonically increases in time. Thus, it is clear that the entropy rate in Eq. (13) is manifestly positive: $dS/dt > 0$. This is the main result of the present work.

Equation (13) with Eqs. (15) and (16) shows how the entropy monotonically increases and therefore is expected to be useful for examining the dynamics of diffusivity fluctuations in the experimental study in Ref. [1].

Based on Eq. (2), we further develop a discussion about the distribution of time being required for characteristic displacement of the RNA-protein particle. The experimental result of the diffusion exponent in Eq. (1) for each cell type is seen to suggest in the random walk picture that large displacements are not significant in contrast, e.g., to Lévy flights [20], meaning the existence of characteristic displacement. Therefore, denoting time being required for characteristic displacement by $\tau$, we assume here that the randomness of $D$ comes from $\tau$ only, (recalling that $\alpha$ is approximately constant for each cell type): $D = D(\tau) \propto \tau^{-\alpha}$, the form of which is due to the dimensional reason. Accordingly, the distribution of $\tau$ is given by $\phi(\tau) \equiv |dD(\tau)/d\tau| P(D)$. In the case when $D/D_0$ is so small [1], for which $\tau$ is large, the distribution decays as a power law, $\phi(\tau) \sim \tau^{-1-\alpha}$. Remarkably, in Ref. [21], where anomalous diffusion of tracer particles in actin filament network has experimentally been observed, such a power-law distribution has been reported for the cage time of the particles (i.e., time required in order for the particles to remain trapped within a localized "cage" of the filaments). Therefore, the power-law nature of $\phi(\tau)$ has an explicit connection with this distribution, since $\tau$ and the local microstructures of the cytoskeletal network of cells studied in Ref. [1] are expected to play roles similar to the cage time and the localized cages of the actin filaments, respectively.

Regarding the entropy in Eq. (4), we make a comment on a formal analogy with the thermodynamic relation concerning temperature. Let us consider the thermodynamic-like situation. There, $D_0$ is supposed to change, the time scale of which should be much larger than the dynamical one. The derivative of the entropy with the distribution in Eq. (2) with respect to $D_0$ is found to be given by $\partial S / \partial D_0 = 1/D_0$. Like the Einstein relation [22], we assume that the diffusivity is proportional to local temperature in a given block. This leads to $D_0 \propto T$ with $T$ being the average value of local temperature over the local blocks, which slowly changes in the present situation. Therefore, we have the following relation: $\partial \widetilde{S} / \partial D_0 = 1/T$, where $S$ has been multiplied by a positive proportional coefficient of $D_0$ and such an entropy is written by $\widetilde{S}$. [It is noted that the maximum entropy principle based on $\widetilde{S}$ also leads to the distribution in Eq. (2) after the redefinition of the Lagrange multipliers.] At the statistical level, this relation has a formal analogy with the thermodynamic relation concerning temperature, if $\widetilde{S}$ and $D_0$ are regarded as the analogs of the "thermodynamic entropy" and the "internal energy" in a given block, respectively.

In conclusion, we have developed the entropic approach for deriving the exponential distribution of diffusivity fluctuations of RNA-protein particle over cytoplasm of *Escherichia coli* cell as well as *Saccharomyces cerevisiae* cell. Then, we have shown how the time derivative of the entropy based on the diffusing diffusivity equation becomes positive. The results mean that the diffusing diffusivity approach is supportive for the statement made in Ref. [1], i.e., the exponential fluctuation as the maximal entropy distribution. We have then found that the distribution of time being required for characteristic displacement of the RNA-protein particle follows a power law. We have also made a comment on a formal analogy with the thermodynamic relation associated with temperature. It is therefore of obvious importance to further examine the present discussions.